\documentstyle[10pt,epsf,epsfig,hangcaption,xspace,amssymb,amsfonts,amsmath,amsthm,cite,
dp_delphititle,lineno]{dp_delphi}
\makeindex
\def\DpPaperGroup{EP}
\def\DpPaperRef{2003-045}
\def\DpDate{7 August 2003}
\def\DpAuthors{DELPHI Collaboration}
\def\DpSubmit{(Eur. Phys. J. C31 (2003) 139-147)}
\def\DpTitle{{ 
Measurement of the $e^+e^- \rightarrow W^+W^-\gamma$ Cross-section and
Limits on Anomalous Quartic Gauge Couplings with DELPHI
}}
\def\DpComment{ }
\def\DpEMail{ }

\newcommand{\gev}{{\ifmmode \mbox{Ge\kern-0.2exV}
\else Ge\kern-0.2exV\nolinebreak\fi}}
\newcommand{\mev}{{\ifmmode \mbox{Me\kern-0.2exV}
\else Me\kern-0.2exV\nolinebreak\fi}}

\begin{document}
\makeatletter
\newcount\@tempcntc
\def\@citex[#1]#2{\if@filesw\immediate\write\@auxout{\string\citation{#2}}\fi
  \@tempcnta\z@\@tempcntb\m@ne\def\@citea{}\@cite{\@for\@citeb:=#2\do
    {\@ifundefined
       {b@\@citeb}{\@citeo\@tempcntb\m@ne\@citea\def\@citea{,}{\bf ?}\@warning
       {Citation `\@citeb' on page \thepage \space undefined}}%
    {\setbox\z@\hbox{\global\@tempcntc0\csname b@\@citeb\endcsname\relax}%
     \ifnum\@tempcntc=\z@ \@citeo\@tempcntb\m@ne
       \@citea\def\@citea{,}\hbox{\csname b@\@citeb\endcsname}%
     \else
      \advance\@tempcntb\@ne
      \ifnum\@tempcntb=\@tempcntc
      \else\advance\@tempcntb\m@ne\@citeo
      \@tempcnta\@tempcntc\@tempcntb\@tempcntc\fi\fi}}\@citeo}{#1}}
\def\@citeo{\ifnum\@tempcnta>\@tempcntb\else\@citea\def\@citea{,}%
  \ifnum\@tempcnta=\@tempcntb\the\@tempcnta\else
   {\advance\@tempcnta\@ne\ifnum\@tempcnta=\@tempcntb \else \def\@citea{--}\fi
    \advance\@tempcnta\m@ne\the\@tempcnta\@citea\the\@tempcntb}\fi\fi}
 
\makeatother
\begin{titlepage}
\pagenumbering{roman}
\CERNpreprint{\DpPaperGroup}{\DpPaperRef} 
\date{{\small\DpDate}} 
\title{\DpTitle} 
\address{\DpAuthors} 
\begin{shortabs} 
\noindent
$W^+W^-\gamma$ production at LEP2 is studied using data
collected with the DELPHI detector
at centre-of-mass energies between 189 GeV and 
209~GeV, correspon\-ding to an integrated luminosity of about 
600~pb$^{-1}$.
Cross-sections are measured for the production of $W^+W^-$ with a
hard, central and isolated photon in the final state, 
and are found to be compatible with the Standard Model prediction.
The photon energy spectra are used to derive limits on anomalous 
contributions to the $W^+W^-Z^0\gamma$ and $W^+W^-\gamma\gamma$ vertices.
\end{shortabs}
\vfill
\begin{center}
\DpSubmit \ \\ 
\DpComment \ \\
\DpEMail \ \\
\end{center}
\vfill
\clearpage
\headsep 10.0pt
\addtolength{\textheight}{10mm}
\addtolength{\footskip}{-5mm}
\begingroup
%
\newcommand{\DpName}[2]{\hbox{#1$^{\ref{#2}}$},\hfill}
\newcommand{\DpNameTwo}[3]{\hbox{#1$^{\ref{#2},\ref{#3}}$},\hfill}
\newcommand{\DpNameThree}[4]{\hbox{#1$^{\ref{#2},\ref{#3},\ref{#4}}$},\hfill}
\newskip\Bigfill \Bigfill = 0pt plus 1000fill
\newcommand{\DpNameLast}[2]{\hbox{#1$^{\ref{#2}}$}\hspace{\Bigfill}}
%
\footnotesize
\noindent
\DpName{J.Abdallah}{LPNHE}
\DpName{P.Abreu}{LIP}
\DpName{W.Adam}{VIENNA}
\DpName{P.Adzic}{DEMOKRITOS}
\DpName{T.Albrecht}{KARLSRUHE}
\DpName{T.Alderweireld}{AIM}
\DpName{R.Alemany-Fernandez}{CERN}
\DpName{T.Allmendinger}{KARLSRUHE}
\DpName{P.P.Allport}{LIVERPOOL}
\DpName{U.Amaldi}{MILANO2}
% ...
\normalsize
\endgroup
\newpage
\titlefoot{Department of Physics and Astronomy, Iowa State
     University, Ames IA 50011-3160, USA
    \label{AMES}}
\titlefoot{Physics Department, Universiteit Antwerpen,
     Universiteitsplein 1, B-2610 Antwerpen, Belgium \\
     \indent~~and IIHE, ULB-VUB,
     Pleinlaan 2, B-1050 Brussels, Belgium \\
     \indent~~and Facult\'e des Sciences,
     Univ. de l'Etat Mons, Av. Maistriau 19, B-7000 Mons, Belgium
    \label{AIM}}
\titlefoot{Physics Laboratory, University of Athens, Solonos Str.
     104, GR-10680 Athens, Greece
    \label{ATHENS}}
\titlefoot{Department of Physics, University of Bergen,
     All\'egaten 55, NO-5007 Bergen, Norway
    \label{BERGEN}}
\titlefoot{Dipartimento di Fisica, Universit\`a di Bologna and INFN,
     Via Irnerio 46, IT-40126 Bologna, Italy
    \label{BOLOGNA}}
\titlefoot{Centro Brasileiro de Pesquisas F\'{\i}sicas, rua Xavier Sigaud 150,
     BR-22290 Rio de Janeiro, Brazil \\
     \indent~~and Depto. de F\'{\i}sica, Pont. Univ. Cat\'olica,
     C.P. 38071 BR-22453 Rio de Janeiro, Brazil \\
     \indent~~and Inst. de F\'{\i}sica, Univ. Estadual do Rio de Janeiro,
     rua S\~{a}o Francisco Xavier 524, Rio de Janeiro, Brazil
    \label{BRASIL}}
\titlefoot{Coll\`ege de France, Lab. de Physique Corpusculaire, IN2P3-CNRS,
     FR-75231 Paris Cedex 05, France
    \label{CDF}}
\titlefoot{CERN, CH-1211 Geneva 23, Switzerland
    \label{CERN}}
\titlefoot{Institut de Recherches Subatomiques, IN2P3 - CNRS/ULP - BP20,
     FR-67037 Strasbourg Cedex, France
    \label{CRN}}
\titlefoot{Now at DESY-Zeuthen, Platanenallee 6, D-15735 Zeuthen, Germany
    \label{DESY}}
\titlefoot{Institute of Nuclear Physics, N.C.S.R. Demokritos,
     P.O. Box 60228, GR-15310 Athens, Greece
    \label{DEMOKRITOS}}
\titlefoot{FZU, Inst. of Phys. of the C.A.S. High Energy Physics Division,
     Na Slovance 2, CZ-180 40, Praha 8, Czech Republic
    \label{FZU}}
\titlefoot{Dipartimento di Fisica, Universit\`a di Genova and INFN,
     Via Dodecaneso 33, IT-16146 Genova, Italy
    \label{GENOVA}}
\titlefoot{Institut des Sciences Nucl\'eaires, IN2P3-CNRS, Universit\'e
     de Grenoble 1, FR-38026 Grenoble Cedex, France
    \label{GRENOBLE}}
\titlefoot{Helsinki Institute of Physics, P.O. Box 64,
     FIN-00014 University of Helsinki, Finland
    \label{HELSINKI}}
\titlefoot{Joint Institute for Nuclear Research, Dubna, Head Post
     Office, P.O. Box 79, RU-101 000 Moscow, Russian Federation
    \label{JINR}}
\titlefoot{Institut f\"ur Experimentelle Kernphysik,
     Universit\"at Karlsruhe, Postfach 6980, DE-76128 Karlsruhe,
     Germany
    \label{KARLSRUHE}}
\titlefoot{Institute of Nuclear Physics,Ul. Kawiory 26a,
     PL-30055 Krakow, Poland
    \label{KRAKOW1}}
\titlefoot{Faculty of Physics and Nuclear Techniques, University of Mining
     and Metallurgy, PL-30055 Krakow, Poland
    \label{KRAKOW2}}
\titlefoot{Universit\'e de Paris-Sud, Lab. de l'Acc\'el\'erateur
     Lin\'eaire, IN2P3-CNRS, B\^{a}t. 200, FR-91405 Orsay Cedex, France
    \label{LAL}}
\titlefoot{School of Physics and Chemistry, University of Lancaster,
     Lancaster LA1 4YB, UK
    \label{LANCASTER}}
\titlefoot{LIP, IST, FCUL - Av. Elias Garcia, 14-$1^{o}$,
     PT-1000 Lisboa Codex, Portugal
    \label{LIP}}
\titlefoot{Department of Physics, University of Liverpool, P.O.
     Box 147, Liverpool L69 3BX, UK
    \label{LIVERPOOL}}
\titlefoot{Dept. of Physics and Astronomy, Kelvin Building,
     University of Glasgow, Glasgow G12 8QQ
    \label{GLASGOW}}
\titlefoot{LPNHE, IN2P3-CNRS, Univ.~Paris VI et VII, Tour 33 (RdC),
     4 place Jussieu, FR-75252 Paris Cedex 05, France
    \label{LPNHE}}
\titlefoot{Department of Physics, University of Lund,
     S\"olvegatan 14, SE-223 63 Lund, Sweden
    \label{LUND}}
\titlefoot{Universit\'e Claude Bernard de Lyon, IPNL, IN2P3-CNRS,
     FR-69622 Villeurbanne Cedex, France
    \label{LYON}}
\titlefoot{Dipartimento di Fisica, Universit\`a di Milano and INFN-MILANO,
     Via Celoria 16, IT-20133 Milan, Italy
    \label{MILANO}}
\titlefoot{Dipartimento di Fisica, Univ. di Milano-Bicocca and
     INFN-MILANO, Piazza della Scienza 2, IT-20126 Milan, Italy
    \label{MILANO2}}
\titlefoot{IPNP of MFF, Charles Univ., Areal MFF,
     V Holesovickach 2, CZ-180 00, Praha 8, Czech Republic
    \label{NC}}
\titlefoot{NIKHEF, Postbus 41882, NL-1009 DB
     Amsterdam, The Netherlands
    \label{NIKHEF}}
\titlefoot{National Technical University, Physics Department,
     Zografou Campus, GR-15773 Athens, Greece
    \label{NTU-ATHENS}}
\titlefoot{Physics Department, University of Oslo, Blindern,
     NO-0316 Oslo, Norway
    \label{OSLO}}
\titlefoot{Dpto. Fisica, Univ. Oviedo, Avda. Calvo Sotelo
     s/n, ES-33007 Oviedo, Spain
    \label{OVIEDO}}
\titlefoot{Department of Physics, University of Oxford,
     Keble Road, Oxford OX1 3RH, UK
    \label{OXFORD}}
\titlefoot{Dipartimento di Fisica, Universit\`a di Padova and
     INFN, Via Marzolo 8, IT-35131 Padua, Italy
    \label{PADOVA}}
\titlefoot{Rutherford Appleton Laboratory, Chilton, Didcot
     OX11 OQX, UK
    \label{RAL}}
\titlefoot{Dipartimento di Fisica, Universit\`a di Roma II and
     INFN, Tor Vergata, IT-00173 Rome, Italy
    \label{ROMA2}}
\titlefoot{Dipartimento di Fisica, Universit\`a di Roma III and
     INFN, Via della Vasca Navale 84, IT-00146 Rome, Italy
    \label{ROMA3}}
\titlefoot{DAPNIA/Service de Physique des Particules,
     CEA-Saclay, FR-91191 Gif-sur-Yvette Cedex, France
    \label{SACLAY}}
\titlefoot{Instituto de Fisica de Cantabria (CSIC-UC), Avda.
     los Castros s/n, ES-39006 Santander, Spain
    \label{SANTANDER}}
\titlefoot{Inst. for High Energy Physics, Serpukov
     P.O. Box 35, Protvino, (Moscow Region), Russian Federation
    \label{SERPUKHOV}}
\titlefoot{J. Stefan Institute, Jamova 39, SI-1000 Ljubljana, Slovenia
     and Laboratory for Astroparticle Physics,\\
     \indent~~Nova Gorica Polytechnic, Kostanjeviska 16a, SI-5000 Nova Gorica, Slovenia, \\
     \indent~~and Department of Physics, University of Ljubljana,
     SI-1000 Ljubljana, Slovenia
    \label{SLOVENIJA}}
\titlefoot{Fysikum, Stockholm University,
     Box 6730, SE-113 85 Stockholm, Sweden
    \label{STOCKHOLM}}
\titlefoot{Dipartimento di Fisica Sperimentale, Universit\`a di
     Torino and INFN, Via P. Giuria 1, IT-10125 Turin, Italy
    \label{TORINO}}
\titlefoot{INFN,Sezione di Torino, and Dipartimento di Fisica Teorica,
     Universit\`a di Torino, Via P. Giuria 1,\\
     \indent~~IT-10125 Turin, Italy
    \label{TORINOTH}}
\titlefoot{Dipartimento di Fisica, Universit\`a di Trieste and
     INFN, Via A. Valerio 2, IT-34127 Trieste, Italy \\
     \indent~~and Istituto di Fisica, Universit\`a di Udine,
     IT-33100 Udine, Italy
    \label{TU}}
\titlefoot{Univ. Federal do Rio de Janeiro, C.P. 68528
     Cidade Univ., Ilha do Fund\~ao
     BR-21945-970 Rio de Janeiro, Brazil
    \label{UFRJ}}
\titlefoot{Department of Radiation Sciences, University of
     Uppsala, P.O. Box 535, SE-751 21 Uppsala, Sweden
    \label{UPPSALA}}
\titlefoot{IFIC, Valencia-CSIC, and D.F.A.M.N., U. de Valencia,
     Avda. Dr. Moliner 50, ES-46100 Burjassot (Valencia), Spain
    \label{VALENCIA}}
\titlefoot{Institut f\"ur Hochenergiephysik, \"Osterr. Akad.
     d. Wissensch., Nikolsdorfergasse 18, AT-1050 Vienna, Austria
    \label{VIENNA}}
\titlefoot{Inst. Nuclear Studies and University of Warsaw, Ul.
     Hoza 69, PL-00681 Warsaw, Poland
    \label{WARSZAWA}}
\titlefoot{Fachbereich Physik, University of Wuppertal, Postfach
     100 127, DE-42097 Wuppertal, Germany \\
\noindent
{$^\dagger$~deceased}
    \label{WUPPERTAL}}
\addtolength{\textheight}{-10mm}
\addtolength{\footskip}{5mm}
\clearpage
\headsep 30.0pt
\end{titlepage}
%
\pagenumbering{arabic} 
\setcounter{footnote}{0} %
\large
\newcommand{\mm}{$\pm$}
\section {Introduction}
\label{sec:Introduction}

$W^+W^-$ production at LEP has been extensively analysed, with both
the total cross-section and the Triple Gauge boson Coupling (TGC) structure
showing good agreement with the Standard Model (SM) predictions~\cite{tgc}.
The high centre-of-mass energy and large data sample available also make 
it possible to study
in detail the events in which a photon is produced together with the 
$W^\pm$ pair.
This paper presents the measurement of the cross-section for
$W^\pm$ pair production with a hard, central and isolated photon 
in the final state.

The Quartic Gauge boson Couplings (QGCs), $W^+W^-Z^0\gamma$ and
$W^+W^-\gamma\gamma$, give rise to 4-fermion$ + \gamma$ final states but their
contribution at LEP2 energies is expected to be 
only about 3 fb in the framework of the SM.
At present energies, these final states result mainly
from initial state radiation (ISR) in $W^+W^-$ production, 
from radiation from the final state charged fermions (FSR) 
or from the intermediate $W^\pm$ boson system (WSR). 
The requirement that the photons be isolated with respect
to the final state charged fermions and the incoming electron/positron beams 
suppresses phase space regions almost entirely dominated by ISR and FSR, and
enhances possible effects from anomalous QGCs.

Deviations from the SM predictions in these final states could imply the
presence of contact interaction contributions to the QGCs, 
signalling new physics whose direct effects are inaccessible at
present energies and could be masked in the TGC measurements. 
The QGC analysis is therefore performed in terms of ``genuine''
quartic gauge couplings, {\it i.e.} excluding those which also give
rise to triple gauge couplings.

Several different parameterizations of the genuine anomalous QGCs have been
given in the literature~\cite{qgc_theory,change}. In this paper, we follow the
analysis of Denner {\it et al}~\cite{change}, which defines five non-SM
Lagrangian operators which conserve electromagnetic gauge invariance and
a custodial $SU(2)_c$ symmetry, and which can give contributions to the
quartic gauge boson vertex. The contributions to the total Lagrangian 
density are:
\vspace{-.5cm}

\begin{eqnarray*}
{\cal{L}}_c & = & \textstyle{-\frac{\pi\alpha}{4\Lambda^2}  a_c F_{\alpha\mu}F^{\alpha\nu}(\overrightarrow{W}_\nu . \overrightarrow{W}^\mu) \, , }\\[0.2cm]
{\cal{L}}_0 & = & \textstyle{-\frac{\pi\alpha}{4\Lambda^2} a_0 F_{\alpha\beta}F^{\alpha\beta}(\overrightarrow{W}_\mu . \overrightarrow{W}^\mu) \, , }\\[0.2cm]
\tilde{\cal{L}}_0 & = & \textstyle{-\frac{\pi\alpha}{4\Lambda^2} \tilde{a}_0 F_{\alpha\beta}\tilde{F}^{\alpha\beta}(\overrightarrow{W}_\mu . \overrightarrow{W}^\mu) \, , }\\[0.2cm]
{\cal{L}}_n & = & \textstyle{-\frac{\pi\alpha}{4\Lambda^2} a_n \epsilon_{ijk}W^{(i)}_{\mu\alpha}W^{(j)}_{\nu}W^{(k)\alpha}F^{\mu\nu} \, , }\\[0.2cm]
\tilde{\cal{L}}_n & = & \textstyle{-\frac{\pi\alpha}{4\Lambda^2} \tilde{a}_n \epsilon_{ijk}W^{(i)}_{\mu\alpha}W^{(j)}_{\nu}W^{(k)\alpha}\tilde{F}^{\mu\nu} \, ,} 
\end{eqnarray*}

\noindent where the field strengths $F_{\mu\nu}$ and 
$\overrightarrow{W}_{\mu\nu}$ are defined by
$F_{\mu\nu}=\partial_\mu A_\nu - \partial_\nu A_\mu$ and 
$\overrightarrow{W}_{\mu\nu}=\partial_\mu \overrightarrow{W}_{\nu} - 
\partial_\nu \overrightarrow{W}_{\mu}$, respectively, and
$2 \tilde{F}_{\mu\nu}=\epsilon_{\mu\nu\rho\sigma}F^{\rho\sigma}  
(\epsilon^{0123}=1)$;
$A_\mu$ is the photon field 
and \\
$\sqrt{2}\overrightarrow{W}_\mu=\left( (W^+ + W^-)_\mu\, , i(W^+ - W^-)_\mu\, , 
\frac{\sqrt{2}}{\cos{\theta_W}}Z_\mu \right)$ is
the triplet of massive gauge bosons. 
The parameter $\Lambda$ has units of energy and represents the scale at 
which new physics would become manifest, and 
 $a_c$, $a_0$, $\tilde{a}_0$, 
$a_n$ and $\tilde{a}_n$ are dimensionless parameters determining the 
separate contributions of each operator.

The first three of these operators contribute to the $W^+W^-\gamma\gamma$ 
coupling and the last two to the $W^+W^-Z^0\gamma$ coupling. 
${\cal{L}}_c$ and ${\cal{L}}_0$ conserve $C$ and $P$, 
$\tilde{\cal{L}}_n$ conserves $CP$ but not the separate symmetries, 
$\tilde{\cal{L}}_0$ conserves only $C$ and 
${\cal{L}}_n$ conserves only $P$.
All these anomalous couplings change not only the total $W^+W^-\gamma$
cross-section but also modify the energy spectra of the observed photons.
In this paper we determine the possible separate contributions 
of each of these operators to $WW\gamma$ production at LEP. 

The results are based on the data collected with the DELPHI detector 
at centre-of-mass energies between 189 and 209~GeV,
corresponding to a total integrated luminosity of about 600~pb$^{-1}$.
Results from the other LEP collaborations on the $W^+W^-\gamma$ final
state and on anomalous QGCs can be found 
in~\cite{lep_qgc_l3,lep_qgc_opal}.

\section{Data Samples}

The data studied in this paper were collected in the LEP runs of 1998 to 2000.
In 1998, DELPHI collected a total luminosity of 154~pb$^{-1}$ at 189~GeV, and
in 1999 integrated luminosities of 
26, 77, 84 and 41~pb$^{-1}$ were recorded at 192, 196, 200 and 202~GeV,
respectively.
In 2000, the centre-of-mass energies ranged from 200 to 209~GeV;  
for part of the year, DELPHI suffered from 
a problem in one of the 12 sectors of its main tracking device (the TPC) 
and, although 
the effect on the present analysis is small, these data were analysed 
separately to isolate any systematic difference. The 2000 data with the 
TPC
fully operational correspond to 161~pb$^{-1}$, and the second set of data 
to 58~pb$^{-1}$, both with average centre-of-mass energies of 206~GeV.

The DELPHI apparatus and performance are described in detail in 
\cite{delphi1,delphi2}. 
The tracking system of DELPHI consisted of a Time Projection Chamber (TPC) 
and a Vertex Detector (VD), 
and was
supplemented by extra tracking detectors, the Inner and Outer Detectors 
in the barrel region, and two Forward Chambers.
It was embedded in a magnetic field of 1.2 T, aligned parallel to the 
beam axis.
The electromagnetic calorimetry consisted of
the High density Projection Chamber (HPC) in the barrel region, the Forward 
Electromagnetic Calorimeter (FEMC) and the Small angle Tile Calorimeter in the
forward regions. 
The regions between the HPC and the FEMC and between HPC modules 
were instrumented by scintillators fitted with lead 
converters so that photons could also be tagged there. 
The hadronic calorimeter covered 98\% of the total solid angle and the 
whole detector was surrounded by muon drift chambers. 
The major hardware change with respect to the description in 
\cite{delphi2}
was the inclusion of the Very Forward Tracker \cite{vft} which extended the 
coverage of the Vertex Detector
down to a polar angle of 11$^{\circ}$. 
Together with new tracking algorithms and alignment 
and calibration procedures, this led to an improved track
reconstruction efficiency in the forward regions of DELPHI.
The tracking algorithms for the barrel part of DELPHI were also
changed to recuperate efficiency in the damaged TPC sector.

The final states considered as $W^+W^-\gamma$ candidates were 
$q\bar{q}q\bar{q}\gamma$ and $q\bar{q}l\nu\gamma$, where $q$ re\-presents 
a 
quark jet and $l \equiv e, \mu, \tau$.
Events corresponding to SM processes with hadronic final states were fully
simulated for the separate data samples. All the
4-fermion final states (including neutral and charged currents) were
generated with the setup described in~\cite{wphact_delphi}, based on 
the WPHACT~\cite{wphact} generator. 
WPHACT was interfaced with YFSWW~\cite{yfsww} to 
include radiative corrections in the Double Pole Approximation 
(DPA) 
approach and to perform ISR corrections 
(including also ISR/WSR interference -- of special relevance for the final
states studied in this paper), while 
PYTHIA~\cite{pythia} modelled the FSR for quarks and
TAUOLA~\cite{tauola} and  PHOTOS~\cite{photos}
modelled the FSR for the charged leptons.
The $q\bar{q}(\gamma)$ final
states were generated with KK2f~\cite{kk2f}. 
For all signal and background processes, the jet fragmentation and 
hadronization was simulated according to the DELPHI tuned JETSET/PYTHIA 
model \cite{tuning}.
All other SM background processes were found to give 
negligible contributions to
the selected samples.

Samples of $W^+W^-\gamma$ with anomalous Quartic Gauge Couplings were 
simulated using weighted events from EEWWG~\cite{EEWWG}. 
This includes a full $O(\alpha)$ calculation of ISR,
WSR and QGC diagrams but not FSR ones; it was interfaced with PYTHIA 
to simulate the
fragmentation, hadronization and FSR from the charged fermions and with
the EXCA\-LIBUR radiator function~\cite{excalibur} to describe collinear 
ISR. 
It was checked that the predictions of EEWWG without anomalous QGCs 
were compatible with the contribution of the $W^+W^-$ production 
tree-level diagrams (CC03)~\cite{cc03} from the WPHACT samples, 
in\-clu\-ding DPA corrections but excluding FSR effects, in the region
studied.
The effect of the collinear ISR is to reduce the effective centre-of-mass 
energy and consequently lower the expected cross-section for visible photons. 
The inclusion of the 
radiative effects from EXCALIBUR in addition to the ISR matrix element in 
EEWWG results in a small double counting of the ISR and of its interference 
with the other contributing processes; however, in the analysis, most of this 
is removed by the use of a subtraction procedure in the event weights, 
as described below.

The EEWWG program is primarily intended to describe anomalous QGC effects for 
on-shell $W^+W^-\gamma$ production: the anomalous signal to be added to 
the SM was defined by applying to each EEWWG event a weight 
$w=w(a_c,a_0,\tilde{a}_0,a_n,\tilde{a}_n)-w(\vec{0})$, defined as
the difference between the matrix element squared calculated with anomalous
couplings $\vec{a} \neq \vec{0}$ and the SM calculation 
($\vec{a} = \vec{0}$)
\footnote{Following \cite{change}, 
$\tilde{a}_0$ and $\tilde{a}_n$ 
were introduced in EEWWG by replacing 
$a_0 \to a_0 + i~\tilde{a}_0$ and $a_n \to a_n + i~\tilde{a}_n$ and
the signs of $a_0/\Lambda^2$
and $a_c/\Lambda^2$ were reversed with respect to the ones in the original
EEWWG code.}. 
Samples of $W^+W^-\gamma$ in all final states were generated with EEWWG
and  fully simulated at centre-of-mass energies of 189, 198 and 206~GeV.
The WPHACT samples were used to define the SM signal to be measured and,
in the analysis of anomalous QGCs,
the extra contribution from EEWWG was added with the weights defined above.
Both the SM and the anomalous QGCs cross-sections obtained in this way were 
found to be compatible with those obtained using RacoonWW \cite{change}.

\section{Event Selection}

The general event reconstruction was based on that used by DELPHI for 
analysis of the process $e^+e^- \rightarrow W^+W^-$~\cite{wwsel}, but
with a less restrictive photon identification in order to enrich the
$W^+W^-\gamma$ sample. 

The reconstruction of photons in DELPHI was done in several steps, starting 
from the showers in the electromagnetic calorimeters.
In the barrel, the procedure described in \cite{delphi2} was followed. Further
``loose'' showers, close to the HPC divisions, were accepted even if they 
failed the transverse shower profile criteria (and also the longitudinal one, 
for showers of energy exceeding 25 GeV). In the forward region, 
all STIC energy deposits with polar angle, $\theta$, with respect to the beam
direction satisfying $3^\circ < \theta < 11^\circ$ were taken to be photon 
candidates\footnote{Energy depositions below 3$^\circ$ were discarded from 
the events, to avoid contamination from off-momentum beam electrons.}, 
while, for $\theta > 11^\circ$,
an algorithm was used to reduce the effects of the shower 
development in the detector material in front of the FEMC:
electromagnetic deposits close in space in the FEMC 
were clustered together and 
the association with reconstructed charged particle tracks was used for 
electron/photon discrimination. 
Care was taken to exclude those tracks which were likely to come from the 
development of showers outside the calorimeter.
Photons with two associated tracks were kept in the ``loose'' selection, but 
the ``tight'' selection required that no VD track elements, 
nor signals from different combinations of other tracking detectors
(depending on the shower polar angle)
be associated with the electromagnetic deposit.
In addition, ``loose'' and ``tight'' photons were required to have a
ratio between the electromagnetic energy 
and the total energy above 90\% in the angular region 
around the cluster defined by $|\Delta{\theta}|<15^\circ$ and 
$|\Delta{\phi}| < \min(15^\circ,6^\circ \cot{\theta_{cluster}})$,
where $\phi$ is the azimuthal angle in the plane perpendicular to
the beam direction and $\theta$ is the polar angle.

The identification of isolated photons in the $W^+W^-\gamma$ samples 
started from the photon candidates defined above and relied on a 
double cone centred around the photon axis, as explained below.
Only isolated photons with energies above 5 GeV were considered.

The total energy inside a cone of 5$^{\circ}$
was associated to the photon, while, to ensure isolation, 
the energy between 5$^{\circ}$ and 15$^{\circ}$ was not permitted 
to exceed 1~GeV.
These criteria were relaxed for tightly identified photons. 
In this case, no further association was done and only the
external cone was considered. The corresponding angle, $\alpha$, 
was varied according to the energy of the 
photon candidate (from 15$^\circ$ down to 3$^{\circ}$ 
for $E_{\gamma}>$ 90~GeV), with the energy inside the cone 
allowed to be reduced proportionally to $\sin\alpha/\sin(15^{\circ})$.
To enrich the FSR sample, identified
muons and electrons coming from the $W^\pm$ 
were excluded from the energy counting within the photon external cone.
All other charged particle tracks (not associated to the photon) 
with momentum greater than 1~GeV/c were required 
to be at least 15$^{\circ}$ away from the photon. 
Although some of the energy of a photon near the electromagnetic 
calorimeter boundaries may be deposited in the HCAL, 
the hadronic energy associated to the photon was required to be below 
5~GeV, and to be less than half of the total photon energy.

The $W^+W^-\gamma$ sample included fully-hadronic ($q\bar{q}q\bar{q}$)
and semi-leptonic ($q\bar{q}e\nu$, $q\bar{q}\mu\nu$ and $q\bar{q}\tau\nu$)
candidate events in which a photon was identified. 
Following the procedures for the 
analysis of $W^+W^-$ events described in \cite{wwsel}, the selection 
of fully-hadronic final states was based on a Neural Network (NN) 
analysis, and that of semi-leptonic final states  
on an Iterative Discriminant Analysis (IDA).
The efficiencies of the selections were of around 80\% for the fully-hadronic 
events and 90\%, 80\% and 65\% for muon, electron and tau semi-leptonic events,
respectively; the purities ranged from 80\% in the fully-hadronic to 99\% in
the muon channel. The background was composed of 75\% $q\bar{q}$ events in 
the fully-hadronic channel and equal amounts of $q\bar{q}$ and 4-fermion 
events in the semi-leptonic channels. 
Since the selection was not tuned specifically for the $W^+W^-\gamma$ process, 
extra cuts were applied to reject background further, as described below.

The measured LEP beam energy values \cite{LEPenergy} were then used in 
kinematic fits,
impo\-sing energy-momentum conservation in the fully-hadronic channel, and
imposing energy-momentum conservation and requiring that the two-jet system 
and the lepton-neutrino system have equal masses in the semi-leptonic 
channels.
In all the fits performed, all the isolated photons were considered to 
come from outside the $W$s, effectively reducing the energy available
for the $W^+W^-$ system.
The $\chi^2$ of the kinematic fits 
had to be below 10 and the corresponding 
fitted quantities were used in the subsequent analysis. 

The maximum energy of the photon was restricted to be below
$(\frac{s-(2M_W)^2}{2\sqrt{s}}-5)$~GeV to select events above the 
$W^+W^-$ threshold.
In the fully-hadronic final state, the output value of the 
$W^+W^-$ Neural Network was required to be larger than 0.70  for all
years.
In the semi-leptonic samples, the lepton and the neutrino reconstructed
by the constrained fit were
required to be isolated in relation to the jets by at least 
10$^{\circ}$ and 5$^{\circ}$, respectively. 

\section{Cross-section Measurement}

\begin{figure}[ht]
\mbox{\epsfig{file=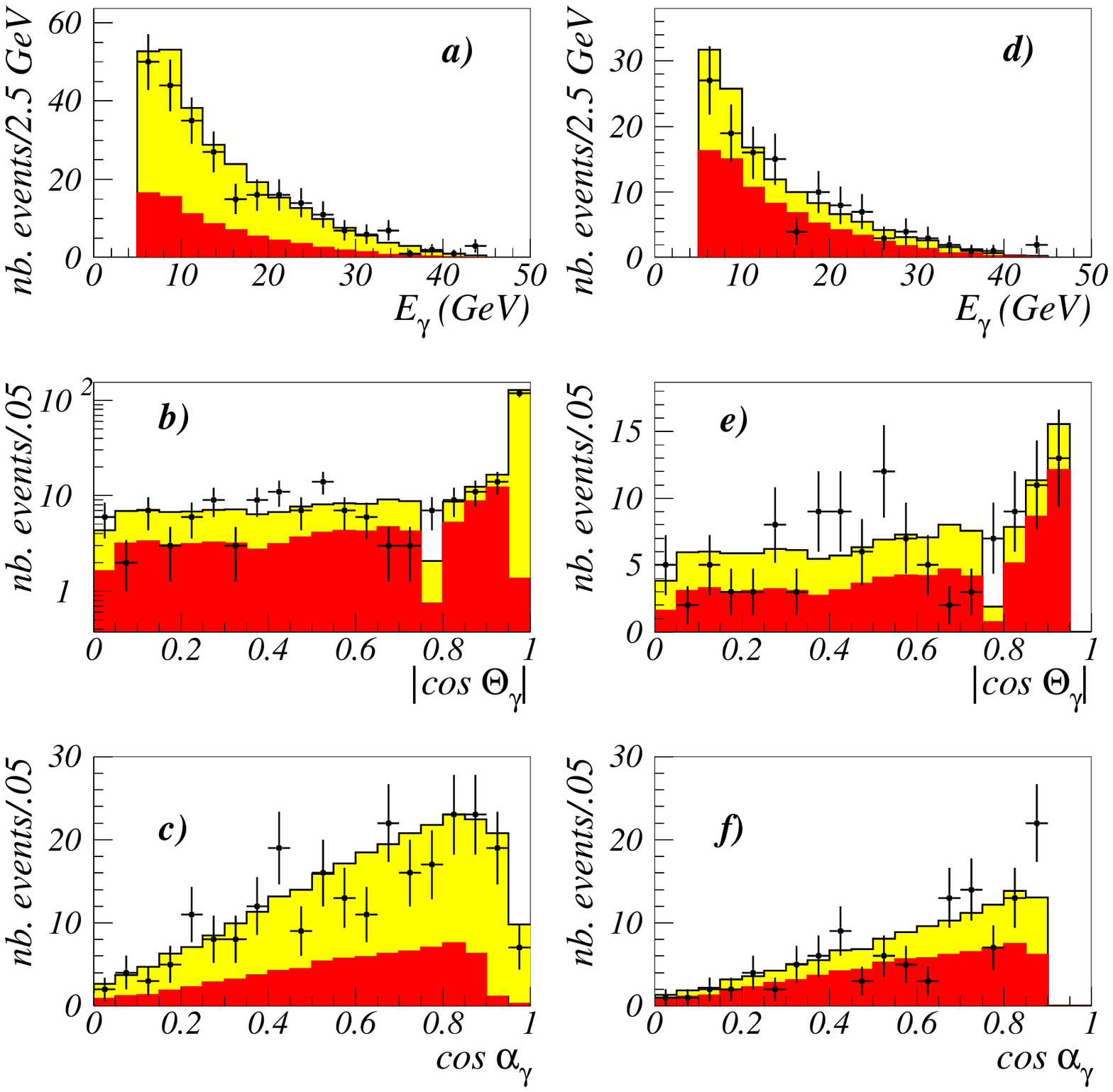,height=0.75\textheight}}
\caption{Distributions of energy (a,d), cosine of the polar angle (b,e) and 
cosine of the isolation angle (c,f) of the photons in all data samples and 
all channels.
a), b) and c) show distributions of the events selected before the
imposition of the signal definition cuts described in the text,
and d), e) and f) the distributions  after imposition of the signal cuts. 
The data (dots)
are compared to the total SM prediction (shaded histogram). The SM 
$W^+W^-\gamma$ signal, generated within the phase space cuts,
is shown in the darker histogram.
In c) and f) events with isolation angle above 90$^\circ$ are not displayed.
}
\label{fig:angles}
\end{figure}

The numbers of events selected as $W^+W^-\gamma$ candidates, according to the 
criteria des\-cribed above, are shown in table~\ref{tab:events} (lines 
labelled ``w/o PS'')
for each channel, dividing the data into three samples according to the
year in which they were collected. 
The distributions of the photon 
energy, polar angle and isolation angle with respect to the reconstructed
jets and leptons are shown in figure~\ref{fig:angles}(a-c), for these events. 
The agreement between data and simulation for the angular variables was
checked in large samples of $q\bar{q}\gamma$ events, corresponding to 
radiative returns to the $Z^0$ pole, 
and no systematic differences were observed. 

The measurement of the  $W^+W^-\gamma$ cross-section was performed in a 
more restricted phase space region in which the photons fulfilled the 
following criteria:

\begin{itemize}
\item{$|\cos{\theta_{\gamma}}| < 0.95$, where $\theta_{\gamma}$ is the
angle between the photon and the incoming electron beam;}
\item{$\cos{\alpha_{\gamma}} < 0.90$, where $\alpha_{\gamma}$ is the
smallest angle between the photon and the final state charged fermions.}
\end{itemize}
The above cuts were applied in addition to the
previously defined criterion:
\begin{itemize}
\item{$E_{\gamma} > 5$ GeV.}
\end{itemize}

The same cuts were applied to the reconstructed events previously selected
(the isolation angle being defined in relation to identified jets and 
charged leptons) 
and the resul\-ting samples used both in the cross-section determination 
and in the anomalous couplings analysis. 
The numbers of events after application of these cuts are also shown in 
table~\ref{tab:events} (2nd row for each data sample listed), and the 
distributions of the energy and angular variables of the photons in these 
samples are shown 
in figure~\ref{fig:angles}(d-f). Events with isolation angles 
$\alpha_{\gamma} > 90^\circ$ are not shown in \ref{fig:angles}(c) and 
\ref{fig:angles}(f):
after imposing the signal definition cuts there are 4 such events in 
data, with 3.3\mm 0.1 expected from simulation.
There is good agreement between the data and the SM expectations, 
except in two samples in the tau channel, 
where a global deficit of around 2 standard deviations 
with respect to the expectations is observed.

\begin{table}[hp]
\begin{center}
\begin{tabular}{|cc|r|r||r||r|r|}
\hline
\multicolumn{2}{|c|}{Channel}&Data&MC$_{tot}$&4-fermion&$WW\gamma$&eff$\times$BF\\
\hline
\multicolumn{7}{c}{ 1998: $\sqrt{s}\sim$189 GeV, $\cal{L}$=154 pb$^{-1}$} \\
\hline
$q\bar{q}e\nu\gamma$ &(w/o PS)&    6 & 8.3\mm0.2 & 8.0\mm0.2 & 3.0\mm0.1 & \\
$q\bar{q}e\nu\gamma$ &   &    4 & 4.0\mm0.2 & 3.8\mm0.1 & 2.9\mm0.1 
& 5.5\% \\
\hline
$q\bar{q}\mu\nu\gamma$&(w/o PS)&   9 &10.2\mm0.2 &10.1\mm0.2 & 3.5\mm0.1 & 
\\
$q\bar{q}\mu\nu\gamma$&   &   5 & 4.6\mm0.2 & 4.6\mm0.1 & 3.5\mm0.1 
& 6.6\% \\
\hline
$q\bar{q}\tau\nu\gamma$&(w/o PS)&  9 &12.4\mm0.3 &10.0\mm0.2 & 3.7\mm0.1 & 
\\
$q\bar{q}\tau\nu\gamma$&   &  1 & 6.4\mm0.3 & 4.9\mm0.2 & 3.6\mm0.1 & 6.8\% \\
\hline
$q\bar{q}q\bar{q}\gamma$&(w/o PS)&25 &31.3\mm0.6 &25.3\mm0.4 & 7.9\mm0.2 & 
\\
$q\bar{q}q\bar{q}\gamma$&   &11 &15.8\mm0.4 &12.6\mm0.2 & 7.6\mm0.2 &14.5\% \\ 
\hline
\hline
sum &(w/o PS)&49 &62.3\mm0.7 &53.3\mm0.5 & 18.0\mm0.3 & 
\\
sum &   &21 &30.7\mm0.5 &25.8\mm0.4 & 17.5\mm0.3 &33.5\% \\ 
\hline
\multicolumn{7}{c}{ 1999: $\sqrt{s}\sim$198 GeV, $\cal{L}$=221 pb$^{-1}$ ($q\bar{q}l\nu\gamma$) and $\cal{L}$=227  pb$^{-1}$ ($q\bar{q}q\bar{q}\gamma$)}\\
\hline
$q\bar{q}e\nu\gamma$ &(w/o PS)&   11 &14.5\mm0.3 &13.6\mm0.2 & 4.8\mm0.1 & 
\\
$q\bar{q}e\nu\gamma$ &   &    5 & 6.7\mm0.2 & 6.2\mm0.2 & 4.7\mm0.1 & 5.6\% \\
\hline
$q\bar{q}\mu\nu\gamma$&(w/o PS)&  15 &16.1\mm0.3 &16.0\mm0.3 & 6.2\mm0.2 & 
\\
$q\bar{q}\mu\nu\gamma$&   &   9 & 7.5\mm0.2 & 7.5\mm0.2 & 6.0\mm0.2 & 7.1\% \\
\hline
$q\bar{q}\tau\nu\gamma$&(w/o PS)& 25 &22.6\mm0.3 &18.6\mm0.3 & 6.5\mm0.2 & 
\\
$q\bar{q}\tau\nu\gamma$&   & 11 &10.9\mm0.2 & 8.6\mm0.2 & 6.3\mm0.2 & 7.5\% \\
\hline
$q\bar{q}q\bar{q}\gamma$&(w/o PS)&56 &52.6\mm0.6 &44.6\mm0.5 &13.8\mm0.3 
&\\
$q\bar{q}q\bar{q}\gamma$&   &30 &25.2\mm0.4 &21.1\mm0.3 &13.4\mm0.3 &15.3\% \\
\hline
\hline
sum &(w/o PS)&107 &105.8\mm0.8 &92.8\mm0.6 & 31.4\mm0.4 & 
\\
sum &   &55 &50.3\mm0.5 &43.3\mm0.4 & 30.4\mm0.4 &35.4\% \\ 
\hline
\multicolumn{7}{c}{ 2000: $\sqrt{s}\sim$206 GeV, $\cal{L}$=199 pb$^{-1}$ ($q\bar{q}l\nu\gamma)$ and $\cal{L}$=219 pb$^{-1}$ ($q\bar{q}q\bar{q}\gamma$)}\\
\hline
$q\bar{q}e\nu\gamma$ &(w/o PS)&   14 &14.9\mm0.3 &14.0\mm0.3 & 4.8\mm0.2 & 
\\
$q\bar{q}e\nu\gamma$ &   &    6 & 6.4\mm0.2 & 6.0\mm0.2 & 4.7\mm0.2 & 5.6\% \\
\hline
$q\bar{q}\mu\nu\gamma$&(w/o PS)&  14 &16.9\mm0.3 &16.8\mm0.3 & 6.1\mm0.2 & 
\\
$q\bar{q}\mu\nu\gamma$&   &   6 & 7.2\mm0.2 & 7.2\mm0.2 & 5.9\mm0.2 & 7.0\% \\
\hline
$q\bar{q}\tau\nu\gamma$&(w/o PS)& 12 &22.1\mm0.4 &18.4\mm0.3 & 6.2\mm0.2 & 
\\
$q\bar{q}\tau\nu\gamma$&   &  5 &10.4\mm0.3 & 8.3\mm0.2 & 6.0\mm0.2 & 7.1\% \\
\hline
$q\bar{q}q\bar{q}\gamma$&(w/o PS)&59 &58.0\mm0.7 &49.3\mm0.5 &14.9\mm0.3 & 
\\
$q\bar{q}q\bar{q}\gamma$&   &29 &27.1\mm0.5 &22.7\mm0.3 &14.4\mm0.3 &15.6\% \\
\hline
\hline
sum &(w/o PS)&99 &111.8\mm0.9 &98.6\mm0.7 & 32.1\mm0.4 & 
\\
sum &   &46 &51.2\mm0.6 &44.1\mm0.4 & 30.9\mm0.4 &35.4\% \\ 
\hline
\end{tabular}
\caption{Number of selected events per channel for each year of data
taking, compared to the expected number of events for the total SM
simulation (MC$_{tot}$).
The numbers corresponding to the contributions of 4-fermion events (and 
specifically of $W^+W^-\gamma$ events) to the total selected simulation sample
are also shown. 
The two lines for each channel show the numbers of events selected as 
$W^+W^-\gamma$ candidates, respectively, 
without (labeled ``w/o PS'') and with the imposition of the 
signal phase space cuts defined in the text.
The efficiency for the signal is shown in the last column for the 
selection within the signal phase space cuts and takes into account the
branching fractions into each channel.
The lower values of luminosity for the semi-leptonic samples reflect 
extra requirements on the detector status.} 
\label{tab:events}
\end{center}
\end{table}

The selections for the four final states 
are exclusive and the efficiencies quoted in 
table~\ref{tab:events} were calculated with respect to the total 
$W^+W^-\gamma$ sample within the signal region, thus including the 
branching fractions to the various decay final states. Note that 
these are not just the SM decay branching ratios of the $W^\pm$, as the 
FSR 
contribution is different for each final state.
The selection purities ($P$) and efficiencies ($\epsilon$) 
vary between channels and also according to the centre-of-mass energies. 
In the final selection, the muon sample has the highest values 
($P \sim$~80\% and $\epsilon \sim$~53\% of the 
$q\bar{q}\mu\nu\gamma$ signal events), followed by
the electron sample, the tau sample and the fully-hadronic sample
with the lowest values ($P \sim$~52\% and $\epsilon \sim$~34\%). 
The samples selected in the tau channel contain substantial contributions
from the other final states, which are taken as signal.

The cross-sections were measured using a likelihood fit to the 
Poissonian probability for observing the 
numbers of events 
shown for each channel in the relevant rows of 
table~\ref{tab:events}:
\begin{itemize}
\item{$\sigma_{WW\gamma}$($\sqrt{s}\sim$ 189 GeV) = 0.19 \mm 0.09 \mm 0.02 pb 
                              (SM: 0.340 \mm 0.017 pb)},
\item{$\sigma_{WW\gamma}$($\sqrt{s}\sim$ 198 GeV) = 0.44 \mm 0.09 \mm 0.03 pb 
                              (SM: 0.385 \mm 0.019 pb)},
\item{$\sigma_{WW\gamma}$($\sqrt{s}\sim$ 206 GeV) = 0.34 \mm 0.09 \mm 0.03 pb 
                              (SM: 0.421 \mm 0.021 pb)},
\end{itemize}
where the first errors are statistical and the second systematic.
The cross-sections obtained with WPHACT/YFSWW for the SM expectations,
in the same phase space region, 
are given with an associated 5\% error \cite{change}.

Although the statistical errors are dominant, conservative 
systematic errors have been estimated from several sources.
The effect of the signal modelling is estimated by varying the photon
distributions (by changing the relative importance of FSR, from the 
expected 25\%-30\% of the signal to 0\% or 50\%,
and of anomalous QGC contributions, within the 
experimentally allowed range determined in the next section), 
and leads to a relative error of 5\% on the global efficiency.
This is complemented with separate contributions of 1.5\% and 3\%  
from the uncertainty in the NN/IDA selections \cite{wwsel} and 
in the photon selection\cite{patricia}, respectively.
A contribution from the modelling of the only significant 
non-4-fermion background, $q\bar{q}(\gamma)$, was obtained conservatively 
by varying the estimated contribution from this channel by $\pm$20\%,
to take into account the uncertainties in the photon production from
fragmentation and in the description of 4-jet observables\cite{4jet+phot}.
The last contribution comes from the finite size of the
simulation samples. 
All the independent contributions were added in quadrature and
the correlations from the errors at different energies were neglected.

\begin{figure}[ht]
\mbox{\epsfig{file=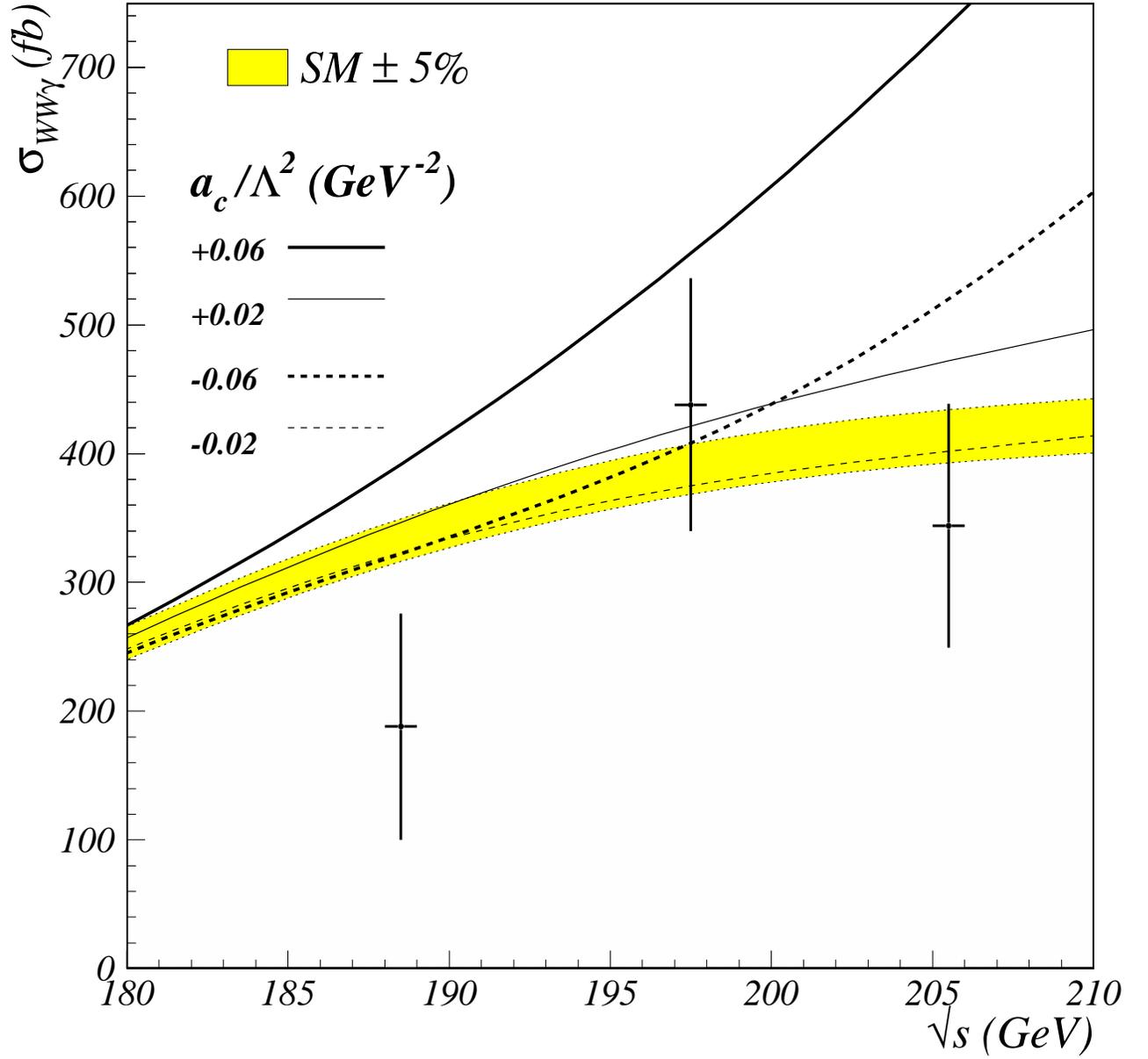,height=0.75\textheight}}
\caption{$W^+W^-\gamma$ cross-section as a function of the centre-of-mass 
energy.
The measured cross-sections (crosses) are compared to the SM prediction 
from WPHACT/YFSWW. The cross-sections obtained with EEWWG for indicative 
values of
the anomalous pa\-rameter $a_c/\Lambda^2$ (in GeV$^{-2}$) are also shown.}
\label{fig:xsec}
\end{figure}

Figure \ref{fig:xsec} shows the comparison between the measured cross-section
and the expected cross-section from WPHACT/YFSWW. 
The changes induced by the presence of anomalous QGCs
for characteristic values of the parameters introduced 
in section~\ref{sec:Introduction}, calculated with EEWWG, 
are also shown.

Another LEP analysis \cite{lep_qgc_l3} has introduced an extra cut 
requiring the two $W^\pm$ bosons to be {\it quasi} on-shell 
($|M_{ff'}-M_W| < 2 \Gamma_W$) and split the 1999 data into two samples. 
In that analysis, the FSR is not considered in the signal. 
We give corresponding results for comparison and to allow for combination 
of LEP results.
Keeping the analysis unchanged but considering the signal defined in this way
leads to a reduction of the signal to between 52\% and 55\% of the original 
value, with slightly higher efficiencies (except for the $q\bar{q}\tau\nu$ 
channel).
The measured cross-sections in this case are:
\begin{itemize}
\item{$\sigma_{WW\gamma}$($\sqrt{s}\sim$189 GeV) = 0.05 \mm 0.08 \mm 0.01 pb 
                              (SM: 0.176 \mm 0.004 pb)},
\item{$\sigma_{WW\gamma}$($\sqrt{s}\sim$195 GeV) = 0.17 \mm 0.12 \mm 0.02 pb 
                              (SM: 0.203 \mm 0.004 pb)},
\item{$\sigma_{WW\gamma}$($\sqrt{s}\sim$200 GeV) = 0.34  \mm 0.12 \mm 0.02 pb 
                              (SM: 0.217 \mm 0.004 pb)},
\item{$\sigma_{WW\gamma}$($\sqrt{s}\sim$206 GeV) = 0.18 \mm 0.08 \mm 0.02 pb 
                              (SM: 0.233 \mm 0.005 pb)}.
\end{itemize}

\noindent 
The theoretical error on these predicted cross-sections is smaller\cite{change}
than that quoted for  the previous selections because only the ISR 
and WSR processes are considered here.

\section{Anomalous Couplings}

\begin{figure}[ht]
\begin{center}
\mbox{\epsfig{file=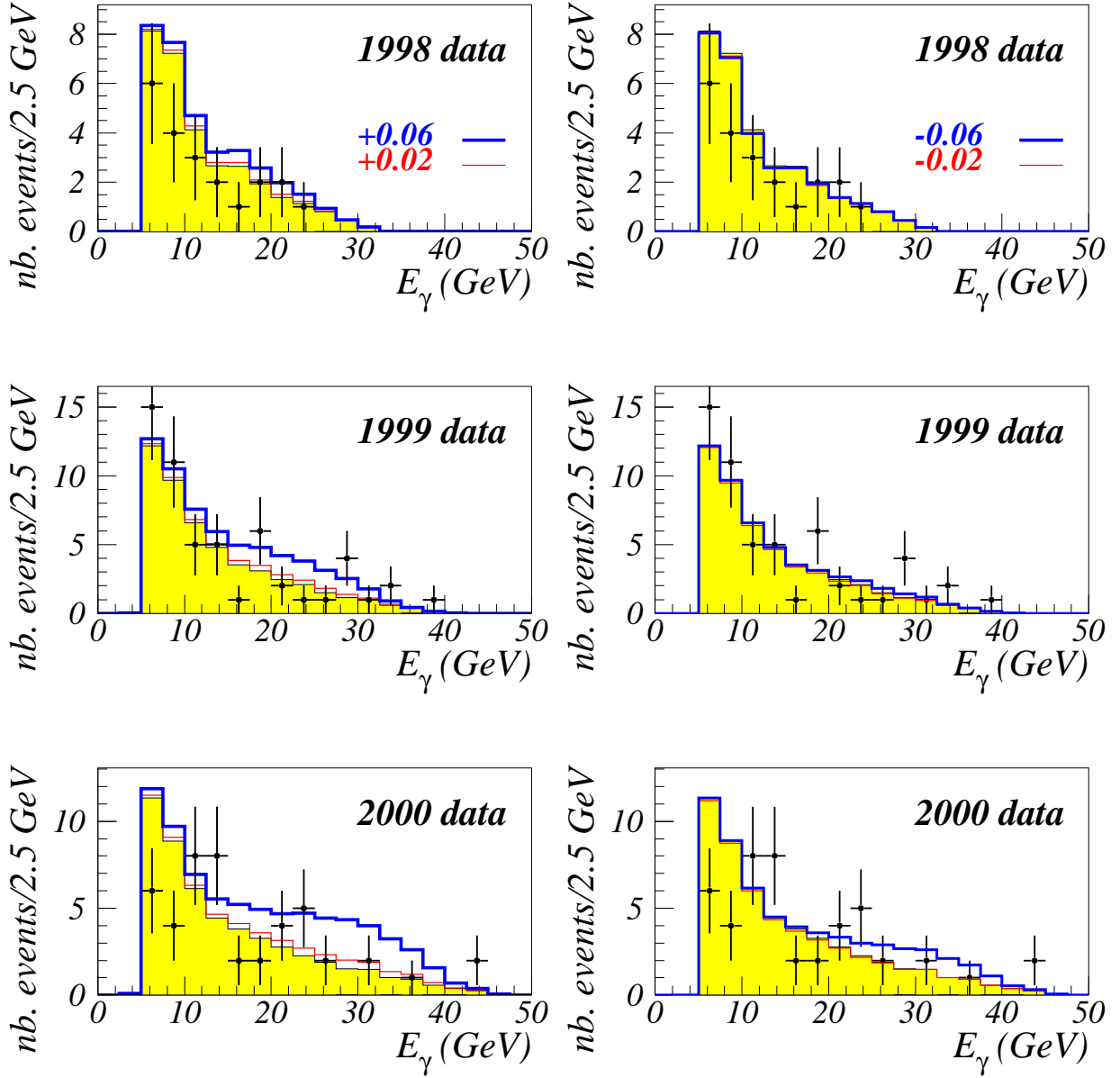,height=0.75\textheight}}
\end{center}
\caption{Effect of anomalous couplings in the photon energy spectra
in each data sample:
data with $\sqrt{s}=$189~GeV (top), data with $\sqrt{s}=$198~GeV (middle)
and data with $\sqrt{s}=$206~GeV (bottom).
The data (dots) are compared to the total SM predictions (shaded 
histogram).
The expected distributions with anomalous QGCs are also shown for 
positive (left) and negative (right) values of $a_c/\Lambda^2$ 
(in GeV$^{-2}$). }
\label{fig:anom}
\end{figure}

The anomalous contributions to the quartic gauge couplings 
were evaluated with EEWWG 
in terms of the parameters $a_c/\Lambda^2$,
$a_0/\Lambda^2$ and $\tilde{a}_0/\Lambda^2$, 
affecting $W^+W^-\gamma\gamma$ vertices, and 
$a_n/\Lambda^2$ and $\tilde{a}_n/\Lambda^2$, 
affecting $W^+W^-Z^0\gamma$ vertices.
These parameters were defined in section~\ref{sec:Introduction}.
The anomalous operators change not only the cross-section but also the photon
energy spectra for the phase space region defined above, and their effects 
are stronger as the centre-of-mass energy increases.
This has been seen in the variation of the total cross-section with 
energy displayed in figure~\ref{fig:xsec}, and is also demonstrated in 
figure~\ref{fig:anom}, which shows the measured and predicted photon 
energy spectra  for the data samples at different energies.
These figures show the SM predictions and also those for four non-zero values 
of $a_c/\Lambda^2$; the distributions predicted for non-zero values of the 
other parameters show the same general behaviour.
The effect on the angular variables is smaller and further reduced by
the selection of central and isolated photons.

A likelihood fit to the photon energy spectra for all the 
individual channels and data sets gives as most probable
values of the QGC parameters 
(in each case setting the values of the others to zero):

\begin{itemize}
\item{$a_c/\Lambda^2=+0.000^{+0.019}_{-0.040}$ GeV$^{-2}$;\\}
\item{$a_0/\Lambda^2=-0.004^{+0.018}_{-0.010}$ GeV$^{-2}$;\\}
\item{$\tilde{a}_0/\Lambda^2=-0.007^{+0.019}_{-0.008}$ GeV$^{-2}$;\\}
\item{$a_n/\Lambda^2=-0.09^{+0.16}_{-0.05}$ GeV$^{-2}$;\\}
\item{$\tilde{a}_n/\Lambda^2=+0.05^{+0.07}_{-0.15}$ GeV$^{-2}$.}
\end{itemize}

\noindent 
Both statistical and systematic uncertainties are included
(the relative contributions are the same as in the cross-section measurement).
At 95\% confidence level, the allowed ranges of the anomalous QGCs are
constrained to be:

\begin{itemize}
\item{$-0.063$ GeV$^{-2} < a_c/\Lambda^2 <$ +0.032 GeV$^{-2}$;}
\item{$-0.020$ GeV$^{-2} < a_0/\Lambda^2 <$ +0.020 GeV$^{-2}$;} 
\item{$-0.020$ GeV$^{-2} < \tilde{a}_0/\Lambda^2 <$ +0.020 GeV$^{-2}$;} 
\item{$-0.18 $ GeV$^{-2} < a_n/\Lambda^2 <$ +0.14  GeV$^{-2}$;} 
\item{$-0.16 $ GeV$^{-2} < \tilde{a}_n/\Lambda^2 <$ +0.17  GeV$^{-2}$.} 
\end{itemize}

The correlation between the different parameters is small and thus these 
results are not substantially changed when multi-parameter fits are 
performed. 
The two parameters which are CP-conserving and affect the 
$W^+W^-\gamma\gamma$ vertex show the largest correlation:
the 95\% upper confidence limit of $a_0/\Lambda^2$ is 0.025~GeV$^{-2}$ 
when $a_0/\Lambda^2$ is fitted together with $a_c/\Lambda^2$.

\section{Conclusions}

About 600 pb$^{-1}$ of LEP2 data, corresponding to centre-of-mass energies
between 189~GeV and 209~GeV, were analysed to study the 
final state $W^+W^-\gamma$, 
where the photon is required to have $E_{\gamma}>$5 GeV,
$|\cos{\theta_{\gamma}}|<$0.95 and to be isolated with respect to the final 
state charged fermions by $\cos{\alpha_{\gamma}}<$0.90.

The cross-sections for $W^\pm$ pair production with a photon in the final state
were found to be in agreement with
the SM prediction and the photon energy spectra were used to test the
presence of anomalous QGCs. 
The data show no evidence for quartic gauge boson couplings.

\subsection*{Acknowledgements}
\vskip 3 mm
 We are greatly indebted to our technical 
collaborators, to the members of the CERN-SL Division for the excellent 
performance of the LEP collider, and to the funding agencies for their
support in building and operating the DELPHI detector.\\
We acknowledge in particular the support of \\
Austrian Federal Ministry of Education, Science and Culture,
GZ 616.364/2-III/2a/98, \\
FNRS--FWO, Flanders Institute to encourage scientific and technological 
research in the industry (IWT), Federal Office for Scientific, Technical
and Cultural affairs (OSTC), Belgium,  \\
FINEP, CNPq, CAPES, FUJB and FAPERJ, Brazil, \\
Czech Ministry of Industry and Trade, GA CR 202/99/1362,\\
Commission of the European Communities (DG XII), \\
Direction des Sciences de la Mati$\grave{\mbox{\rm e}}$re, CEA, France, \\
Bundesministerium f$\ddot{\mbox{\rm u}}$r Bildung, Wissenschaft, Forschung 
und Technologie, Germany,\\
General Secretariat for Research and Technology, Greece, \\
National Science Foundation (NWO) and Foundation for Research on Matter (FOM),
The Netherlands, \\
Norwegian Research Council,  \\
State Committee for Scientific Research, Poland, SPUB-M/CERN/PO3/DZ296/2000,
SPUB-M/CERN/PO3/DZ297/2000 and 2P03B 104 19 and 2P03B 69 23(2002-2004)\\
JNICT--Junta Nacional de Investiga\c{c}\~{a}o Cient\'{\i}fica 
e Tecnol$\acute{\mbox{\rm o}}$gica, Portugal, \\
Vedecka grantova agentura MS SR, Slovakia, Nr. 95/5195/134, \\
Ministry of Science and Technology of the Republic of Slovenia, \\
CICYT, Spain, AEN99-0950 and AEN99-0761,  \\
The Swedish Natural Science Research Council,      \\
Particle Physics and Astronomy Research Council, UK, \\
Department of Energy, USA, DE-FG02-01ER41155, \\
EEC RTN contract HPRN-CT-00292-2002. \\


\newpage
\vspace{-0.5cm}

\vspace{-0.5cm}
\end{document}